\documentstyle[prl,aps,epsfig,floats,amssymb, amsmath,bbm]{revtex}

\begin{document}

\draft

\twocolumn[\hsize\textwidth\columnwidth\hsize\csname@twocolumnfalse%
\endcsname

\title{Broken Ergodicity in a Stochastic Model with Condensation}

\author{Frank Zielen and Andreas Schadschneider} 

\address{Institute for Theoretical Physics, 
University of Cologne,
Z\"ulpicher Stra\ss{}e 77,
D-50937 K\"oln, Germany} 

\maketitle 

\begin{abstract}
  We introduce a variant of the asymmetric random average process with
  continuous state variables where the maximal transport
  is restricted by a cutoff. For periodic boundary
  conditions, we show the existence of a phase transition between a 
  pure high flow phase and a mixed phase, whereby the latter consists of a 
  homogeneous high flow and a condensed low flow substate without
  translation invariance.
  The finite system alternates between these substates which both have
  diverging lifetimes in the thermodynamic limit, so ergodicity is broken
  in the infinite system. However, the scaling behaviour of the lifetimes
  in dependence of the system size is different due to
  different underlying flipping mechanisms. 
\end{abstract}

\pacs{05.40.-a, 
45.70.Vn,
02.50.-r, 
05.60.-k 
}
]


The {\em asymmetric random average process (ARAP)} \cite{KrugJ:asyps,RajeshR:conmm} represents a simple fundamental model for numerous applications, especially in the field of interdisciplinary questions. It has a close relationship to the q model describing force distributions in granular materials \cite{CoppersmithSN:forfbp}, aggregation and fragmentation processes (e.g.\ mass transport problems) \cite{majumdar:diffaggrfrag} and traffic flow theory, in particular a continuous Nagel-Schreckenberg model \cite{krauss:contns}.



The {\em truncated ARAP (TARAP)} studied in this letter, although equipped with rather simple homogeneous short-range interaction and periodic boundary conditions, 
features the surprising behaviour of broken ergodicity, i.e.\@
a state space decomposition into dynamically unconnected subsets. In addition, these substates are totally different: the system may reside either in a homogeneous high flow phase or in a condensed low flow phase where a finite fraction of the total mass is located on a randomly chosen site. While examples of 
infinite aggregation \cite{majumdar:diffaggrfrag} and ergodicity breaking \cite{Evans:ssb_letter,arndt:ssb} are known for 1D nonequilibrium systems, the simultaneous occurrence of these phenomena seems to be undiscussed and relevant for a lot of applications.

The ARAP is defined on a 1D periodic lattice with $L$ sites. 
Each site $i$ carries a non-negative continuous mass variable 
$m_i \!\in\! {\mathbbm R}^{+}_{0}$. In every discrete time step
$t\to t+1$ for each site a random number $r_i \in [0,1]$ is generated
from a time-independent probability density function (pdf) $\phi$, 
sometimes called {\em fraction density}, that may depend on the actual 
configuration $m=(m_1,\ldots,m_L)$. So the universal form can be 
written as $\phi=\phi(r_1,\ldots,m_1,\ldots)$. The fraction $r_i$
determines the amount of mass $r_i m_i$ transported from site $i$ to
site $i+1$. The transport is completely asymmetric, i.e.\ no mass moves
in the opposite direction $i+1\to i$, and we get
\begin{eqnarray} \label{mass_update}
m_i \rightarrow (1-r_i) m_i + r_{i-1} m_{i-1} \;.
\end{eqnarray}
These update rules correspond to a parallel dynamics. Due to the
conservation of the total mass $M=\sum_i m_i$ the density
$\rho=\frac{M}{L}$ is fixed. We would like to add that this so called
stick representation is equivalent to a particle picture
\cite{KrugJ:asyps,RajeshR:conmm} not used in this letter.
 
The simplest version of the ARAP is obtained if we use the state
independent uniform distribution defined by the fraction density
$\phi=1$. We will refer to this system as the {\em free ARAP}. Some
results have been derived for this model so far whereby the exactness of
product measure for $L\rightarrow \infty$ and its form 
\cite{KrugJ:asyps,RajeshR:conmm,CoppersmithSN:forfbp}
\begin{equation} \label{free_mass_prob}
P(m) = \prod_i P(m_i) \;\;\; \text{with} \;\;\;
P(m_i)=\frac{4 m_i}{\rho^2} \;e^{-2m_i/\rho}
\end{equation}
is the most relevant for our work. Eq.~(\ref{free_mass_prob}) has been 
obtained in the context of the {\em q model} \cite{CoppersmithSN:forfbp} 
that shares many properties with the ARAP.
Further results for the ARAP can be found in 
\cite{KrugJ:asyps,RajeshR:conmm,SchuetzG:tradc,rajesh:corrrap,zielenschad:emfs}.

In the ARAP the mass $r_{i}m_{i}$ transferred from site $i$ to site 
$i+1$ is in principle unbounded. This is different in the 
truncated ARAP that we introduce here. In the TARAP all transfers of masses 
$r_{i}m_{i}$ larger than a cutoff $\Delta>0$ are rejected.
We like to focus on the simplest model only, the 
{\em truncated free ARAP}.
The corresponding fraction density is then given by 
$\phi=\prod_i \phi(r_i,m_i)$ with
\begin{equation} \label{phi_delta}
\phi(r_i,m_i) = \left[ 1-R(m_i) \right] \delta(r_i) + 
\Theta\left( R(m_i) - r_i \right) \;,
\end{equation}
where $\Theta$ is the Heaviside step-function and
\begin{equation} \label{r_delta}
R(m_i) \equiv \min\left(1, \frac{\Delta}{m_i} \right)
\end{equation}   
represents the maximum possible fraction. Note that $\phi$ has
become locally state dependent, i.e.\ the pdf depends on the mass $m_i$
at the corresponding site explicitly. 

\begin{figure}[hbt]
\begin{center}
\epsfig{file=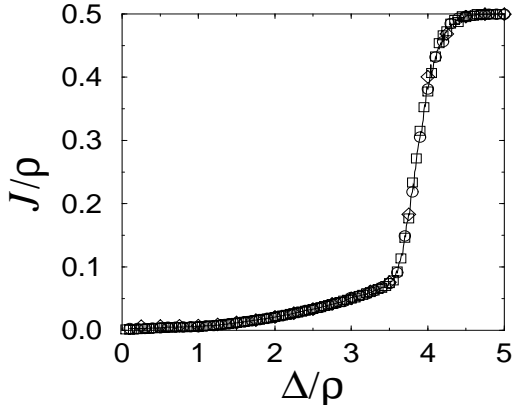, height=0.65\columnwidth, width=0.85\columnwidth}
\caption{Current-cutoff diagram for system size $L=100$ and different 
densities $\rho$ $( \frac{2}{5}=\diamond, \; 1=\circ, \; 2=\Box )$.}
\label{funda1}
\end{center}
\end{figure}

Without loss of generality we set $\rho=1$ for the rest of the letter
because every TARAP defined by $(\rho,\Delta)$ can be mapped onto a
$(1,\frac{\Delta}{\rho})$-system (Fig.~\ref{funda1}). 
Furthermore we introduce the rescaled cutoff 
\begin{equation}
\tilde{\Delta} = 2 L^{-\frac{1}{2}} \Delta 
\end{equation}
which ensures $L$-independence of the critical point.

We begin by investigating the relation between the steady state
current $J$, defined by the average mass transfer per site, and 
the cutoff parameter $\Delta$ (Fig.~\ref{funda1}). 
For $\Delta \rightarrow 0$ the flow vanishes because the transferred 
mass per site is always smaller than $\Delta$ while for 
$\Delta \rightarrow \infty$ the system behaves like a free ARAP and the 
value of $J$ tends to its maximum $J^{\text{max}} \equiv \frac{1}{2}$. 

If we study the process in more detail we discover the system
to exist in two phases whereby the transition point $\tilde{\Delta}_c
= 1$ has been determined numerically and by analytical approximations.

For rescaled cutoffs $\tilde{\Delta}>\tilde{\Delta}_c$ the steady
state mass distribution is nearly identical to the one of the free
ARAP and approximately given by (\ref{free_mass_prob}). The flow is 
independent of $\tilde{\Delta}$ and corresponds to the maximum
current $J^{\text{max}}_{\text{high}} \equiv J^{\text{max}}$. 
Therefore we refer to this parameter range as the {\it high flow phase}.

For rescaled cutoffs smaller than $\tilde{\Delta}_c$ the steady state
of the finite system is a composition of two different substates: the
system can either exist in a high flow state (with properties as
described above) or a state given by a macroscopic condensate, i.e.\ a
finite fraction of the total mass $M$ resides on one randomly chosen site (even in the
thermodynamic limit). So the mass aggregation is proportional to $L$, but not extensive in space.
 The remaining mass is distributed equally 
with an algebraically decaying mass distribution. 
The current of the condensate
state depends on $\tilde{\Delta}$ but does not exceed $J_{\text{low}}^{\text{max}} \equiv
\frac{1}{4}$. Therefore we call this state {\em low flow state}
and denote the parameter region $\tilde{\Delta}<\tilde{\Delta}_c$ as
{\it mixed phase}.

\begin{figure}[hbt]
\begin{center}
\epsfig{file=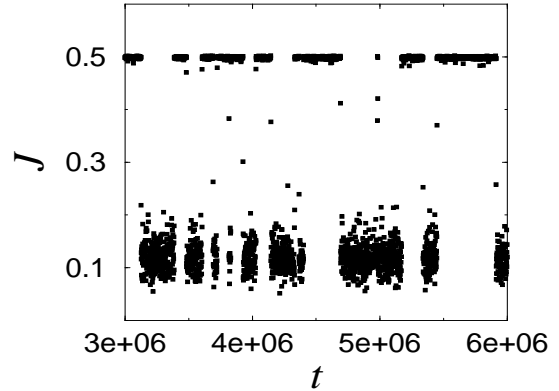, height=0.65\columnwidth, width=0.85\columnwidth}
\end{center}
\caption{Current-time diagram of the mixed flow phase. Parameters used: 
$L=100$ and $\tilde{\Delta}\!=\!0.846$. Each current value is averaged 
over $10^3$ time steps.}
\label{current_time}
\end{figure}

In a finite system the transition probabilities between low and high
flow states are small but nonzero. The system switches
between these states while evolving in time and an alternating
current-time relation is obtained (Fig.~\ref{current_time}).
Note that the switching time between the two states is much smaller
than their lifetimes.

This alternating behaviour is very similar to systems with spontaneously broken symmetry \cite{Evans:ssb_letter,arndt:ssb}. However, in general the flipping is between states of broken symmetry. For example in \cite{Evans:ssb_letter} an ASEP with two particles ($+$) and ($-$) is introduced and in regimes of spontaneously broken symmetry the system switches between states dominated by ($+$) or ($-$) particles.
In case of the TARAP the flipping occurs between the symmetric high flow state which is translation invariant and a low flow state in which this symmetry is broken. 

\begin{figure}[hbt]
\begin{center}
\epsfig{file=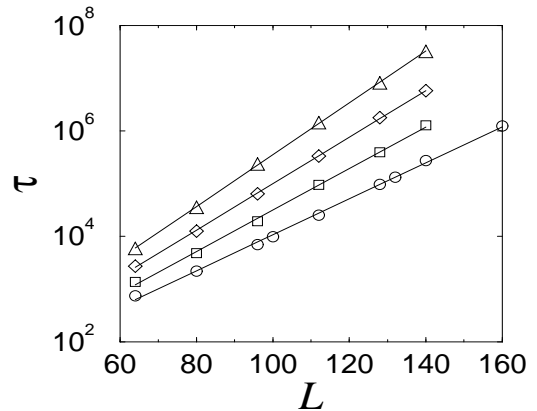, height=0.65\columnwidth, width=0.85\columnwidth}
\end{center}
\caption{Linear-logarithmic plot of the lifetime $\tau_H$ of the
high flow state in dependence of the system size $L$ for several 
rescaled cutoffs $\tilde{\Delta}$ $=$ $0.76(\circ)$, $0.80(\Box)$, 
$0.84(\Diamond)$ and $0.88(\triangle)$.}
\label{lifetime_high}
\end{figure}

\begin{figure}[hbt]
\begin{center}
\epsfig{file=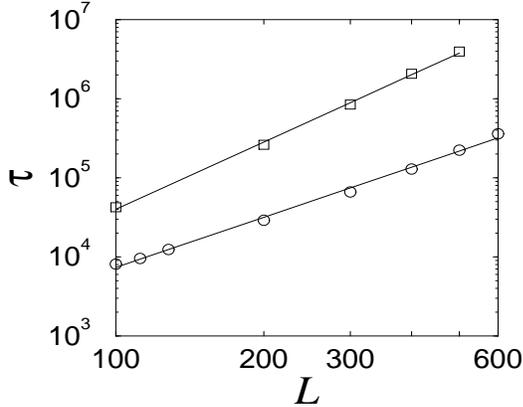, height=0.65\columnwidth, width=0.85\columnwidth}
\end{center}
\caption{Log-log plot of the lifetime $\tau_L$ of the low flow state in 
dependence of the system size $L$ for the rescaled cutoffs 
$\tilde{\Delta}$ $=$ $0.92(\circ)$ and $0.88(\Box)$.}
\label{lifetime_low}
\end{figure}

In the thermodynamic limit the average lifetimes $\tau_H$ and
$\tau_L$ of the high and low flow states diverge (Figs.~\ref{lifetime_high} 
and \ref{lifetime_low}).
This implies that the steady state in the mixed phase is not unique. 
Ergodicity is broken and the steady state can either be in the
{\it high flow state} or the {\it low flow state} depending on
the initial condition. However, the lifetimes of the substates do not scale equally with size $L$ (see next paragraph).

Although Monte Carlo simulations are difficult in the mixed phase
since the lifetimes of the substates are very large
it is safe to say that both the average 
lifetimes of low and high flow states diverge in the thermodynamic
limit.
Fig.~\ref{lifetime_high} indicates that $\tau_H$ increases exponentially
for all $\tilde{\Delta}<\tilde{\Delta}_c$.
Assuming $\tau_H \sim \exp(\alpha(\tilde{\Delta}) L)$ 
we obtain that $\alpha$ is an increasing function of
$\tilde{\Delta}$. 
Note that in the pure high flow phase $\tilde{\Delta}>\tilde{\Delta}_c$
the relation $\tau_H = \infty$ holds for any $L$. 
The corresponding data for $\tau_L$ in the low flow state are better 
fitted algebraically than exponentially (Fig.~\ref{lifetime_low}). 
We see again that the exponents obtained by the assumption $\tau_L \sim
L^{\beta(\tilde{\Delta})}$ increase while moving into the mixed phase,
i.e.\ for $\tilde{\Delta}$ tending to zero.

Although both transitions are driven by fluctuations,
the above scaling behaviours reflect different switching mechanisms. 
While the high $\to$ low transition is based on a collective effect
which involves all sites of the lattice, the opposite transition
involves only the lattice site where the macroscopic condensate is
located. Under these assumptions we have analytically derived approximations that  
reflect qualitatively the lifetime behaviour \cite{fzas:diss}. 

Note that in spontaneous symmetry breaking both transitions obey the same
purely fluctuation driven mechanism yielding the same exponential behaviour
of the lifetimes. A flipping mechanism between two alternating symmetric substates that
is not purely driven by fluctuations is studied in
\cite{loan:alternating}, resulting in a logarithmic lifetime dependence.

The above results are mainly based on Monte Carlo simulations.
We continue now by presenting a few analytical approximations which support 
the numerical findings.
     
Starting point is a closer look at the average 
mass transfer per site (local current)
\begin{equation} \label{delta_function}
J(m) \equiv \left\langle r m \right\rangle_{\phi(r,m)} =
\begin{cases} 
\frac{1}{2} m & \ \ {\rm for\ \ } 0 \leqslant m < \Delta \\
\frac{\Delta^2}{2m} & \ \ {\rm for\ \ } \Delta \leqslant m < \infty 
\end{cases} \;.
\end{equation}
This shows that the average mass shift $J(m)$ tends to 
zero for $m\rightarrow\infty$ and $m\rightarrow 0$. So high (low) 
columns shrink (grow) very slowly and accordingly low flow states, 
resp.\ one-stick configurations, are very stable. 
On the other hand, homogeneous configurations maximize 
the current. We will exemplify this in the following paragraphs.

First we study the low flow state using the approximation
\begin{equation} \label{low_flow_approximation}
\left \langle J(m) \right \rangle_{P_i} = J\left( \langle m 
\rangle_{P_i}\right)\;.
\end{equation}
Here $P_i$ denotes the steady state single-site distribution
of site $i$. Equation (\ref{low_flow_approximation}) holds for stable
distributions $P_i(m) \sim \delta(m-m_i)$ which we assume here.
Introducing the nomenclature $m_i \equiv \langle m \rangle_{P_i}$ we
obtain from the continuity equation in the steady state 
\begin{equation}\label{continuity}
J_{\text{low}} \equiv J(m_i) = J(m_{i+1})\qquad\quad
\text{for\ all\ } i. 
\end{equation}
Restricting now to the
case where only one high column exists at site $j$, i.e. $m_+ \equiv
m_j > \Delta$, the remaining columns are all of the same mass
(see (\ref{continuity})), i.e.\ $m_- \equiv m_i < \Delta$ with 
$i \not= j$. So, using (\ref{delta_function}) and (\ref{continuity}),
the quantities $m_{\pm}$ are related by $m_+ m_- = \Delta^2$ and 
the mass conservation law $\rho L = (L-1) m_- + m_+$.

\begin{figure}[hbt]
\begin{center}
\epsfig{file=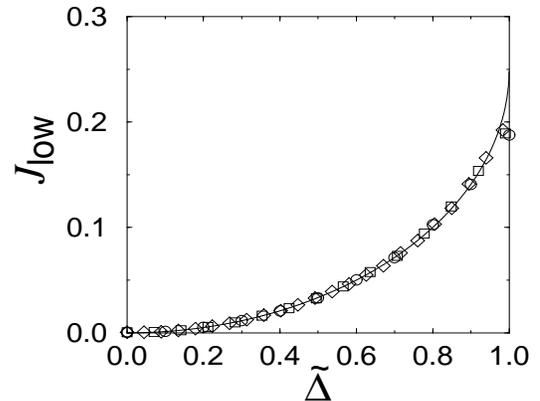, height=0.65\columnwidth, width=0.85\columnwidth}
\end{center}
\caption{Low flow state current in dependence of the rescaled cutoff. 
Comparison of analytical (-) and numerical ($L=100(\circ)$, 
$200(\Box)$, $500(\Diamond)$) data.}
\label{jlowhigh_cutoff}
\end{figure}

Equipped with these relations we are able to compute $m_{\pm}$ and
finally the current in the low flow state $J_{\text{low}}$. Assuming
$L \gg 1$ our calculations lead to the formula
\begin{equation} \label{jlow_ana}
J_{\text{low}} = J_{\text{low}}^{\text{max}} \left\{ 1 - 
\sqrt{1 - \tilde{\Delta}^2}\right\} \;.
\end{equation}
So the current of the low flow state increases from zero to
$J_{\text{low}}^{\text{max}}$ for $0 \!<\! \tilde{\Delta}
\!\leqslant\! \tilde{\Delta}_c$.  For rescaled cutoffs larger than
$\tilde{\Delta}_c$ relation (\ref{jlow_ana}) is not defined and the
one-stick model is not valid anymore. Thus we have derived an upper bound 
for the occurrence of the low flow state in accordance with the
simulations. As Fig.~\ref{jlowhigh_cutoff} shows the analytical result
agrees very well with Monte Carlo data.

By extending the calculations above to $N$ mass aggregates \cite{fzas:diss} the
validity range in terms of $\tilde{\Delta}$ shrinks and we obtain
smaller critical cutoffs $\tilde{\Delta}_c$. Furthermore the flux
increases. This indicates that the mixed phase is described best by
the one-condensate picture, especially in the vicinity of the critical
point. This is confirmed by numerical investigations that
have shown that configurations with two or more aggregates are not
stable. 
The pre-condensates
struggle for masses until only one stick is left.
This mechanism of \emph{competition} is different to \emph{coarsening} where aggregates emerge at several sites and then coalesce \cite{majumdar:diffaggrfrag}.

Focussing on the high flow state, resp.\ phase, 
every site carries the same mass $\rho$ on average. 
For large $L$ we approximate, motivated by the numerical results, 
the mass distribution by the product measure solution of the free 
ARAP (\ref{free_mass_prob}) and calculate by the help of 
(\ref{delta_function}) the high flow state current
\begin{equation} \label{jhigh_ana}
J_{\text{high}}= J^{\text{max}}_{\text{high}} \left\{ 
1 - (1+L^{\frac{1}{2}} \tilde{\Delta}) 
e^{- L^{\frac{1}{2}} \tilde{\Delta}} \right\} 
\underset{L \gg 1}{\longrightarrow} J^{\text{max}}_{\text{high}}
\end{equation}
for all rescaled cutoffs. So the high flow state exists both
in the mixed phase and in the high flow phase. Equation
(\ref{jhigh_ana}) also predicts that the current in the high flow
state differs from $J^{\text{max}}_{\text{high}}$ only for small
absolute cutoffs $\Delta$ in finite systems. We have verified this by
Monte Carlo simulations although it is no longer possible to
distinguish the states by their flows which are nearly identical in
this parameter range. So we used the appearance of a macroscopic
condensate as a criterion.

A model that is strongly related to the TARAP
is studied in \cite{krauss:contns,nageltgf01} where 
a similar kind of phase separation as in the mixed domain has been 
observed. The underlying process is the continuum limit (Krauss model) 
of the Nagel-Schreckenberg cellular automaton model \cite{review}
that is defined by continuous velocities and spatial coordinates. 
Although this traffic model is given by more complex dynamics than 
the TARAP, both processes have a common feature:
moves may be rejected if a uniformly distributed random variable
exceeds a given threshold. So the TARAP can be viewed as a 
toy model that catches some of the fundamental physics behind 
the Krauss model.
We have also rewritten a simplified version of the Krauss model in terms of fraction densities to point out the similarities with the TARAP explicitly \cite{fzas:diss}.


In the Krauss model an additional congested phase has been observed \cite{krauss:contns}.
An analogous phase,
corresponding to a pure condensed phase, could be expected for the
TARAP in the limit of small cutoffs. However, in that regime the
lifetime of the low flow state is much larger than the corresponding 
lifetime of the high flow state, in particular for small systems.
Therefore, in Monte Carlo simulations it is difficult to distinguish between a pure condensed phase and the mixed regime.
Furthermore, our analytical
calculations have not shown any evidence for a transition
point separating a (new) pure low flow phase and the mixed phase.
   

What are the essential ingredients of the TARAP leading to the observed phenomena?
First \emph{truncation} and an \emph{unbounded local state space} allow for an unlimited condensation. The formation of infinite aggregates can also be found in a model of aggregation and fragmentation \cite{majumdar:diffaggrfrag,oloan:brm} or a special zero-range process \cite{evans:zrp}, both equipped with unbounded but discrete state variables. However, in these examples the condensed phase is unique and not associated with a high flow counterpart, i.e.\@ high flow phase and congested phase are separated by a phase transition. The same is true for a discretized version of the TARAP that we have investigated \cite{fzas:diss}.

Therefore we believe that in case of the TARAP the \emph{continuous state variables} allow for the coexistence of high and low flow state in the mixed phase, yielding nonsymmetric ergodicity breaking in the \emph{thermodynamic limit}. The TARAP is free of an intrinsic mass scale in contrast to models defined on integer state space, equipped with a smallest mass unit $1$. Thus, in the TARAP we obtain a trivial dependence on the density $\rho$ which is also reflected in the scaling law $(\rho,\Delta) \leftrightarrow (1,\Delta/\rho)$. However, in discrete systems physics may change in the low density regime, e.g.\@ an ARAP with total mass $M\!=\!1$ is nothing else than a one-particle ASEP. This difference may indicate why in the continuous variant the homogeneous high flow state can persist over the whole parameter range, resulting in the coexistence phenomenon.

We would like to thank Joachim Krug, Kavita Jain and Martin Evans for
interesting und helpful discussions.

\bibliographystyle{unsrt}

\end{document}